# EVIDENCE FOR RESIDUAL MATERIAL IN ACCRETION DISK GAPS: CO FUNDAMENTAL EMISSION FROM THE T TAURI SPECTROSCOPIC BINARY DQ TAU

John S. Carr[1]

Naval Research Laboratory, Code 7213, Washington, DC 20375; carr@mriga.nrl.navy.mil

Robert D. Mathieu[1]

Department of Astronomy, University of Wisconsin, Madison, WI 53706; mathieu@astro.wisc.edu

and

Joan R. Najita[1]

National Optical Astronomy Observatories, 950 N. Cherry Ave., Tucson, AZ 85719;

najita@noao.edu

[1] Visiting Astronomer at the Infrared Telescope Facility which is operated by the University of Hawaii under contract to the National Aeronautics and Space Administration

ABSTRACT

We present the discovery of CO fundamental ro-vibrational emission from the classical T Tauri spectroscopic binary DQ Tau. The high-resolution infrared echelle spectra reveal emission lines from both the $v=1$ and $v=2$ vibrational levels with line widths of roughly 70 km s$^{-1}$. The average CO excitation temperature is approximately 1200 K. We model the spectra as arising from gas in Keplerian rotation about the center-of-mass of the binary. The disk model requires gas with an average surface density of 5 x 10$^{-4}$ g cm$^{-2}$ that extends outward to 0.5 ± 0.1 AU and inward to at least 0.1 AU from the center-of-mass. The radial extent for the emitting gas is close to the predicted size of the gap in the DQ Tau accretion disk that is expected to be dynamically cleared by the binary. We interpret these results, and previous modeling of DQ Tau's spectral energy distribution, as evidence for a small amount (~ 10$^{-10}$ M$_\odot$) of diffuse material residing within the optically-thin disk gap. Thus dynamical clearing has not been completely efficient in the DQ Tau binary. We suggest that the material is associated with a flow from the circumbinary disk which feeds the ongoing accretion at the stellar surfaces.

Subject Headings: accretion, accretion disks --- binaries: spectroscopic --- stars: pre-main-sequence

1. INTRODUCTION

Among the classical T Tauri stars (CTTS), close binary systems offer interesting insights into the physics of disk accretion, physics which also has an important bearing on the formation of giant planets within protoplanetary disks. Theory predicts that the tidal interaction of a young binary system with its accretion disk will clear away disk material near the radius of the binary separation, creating a gap in the disk that separates an outer circumbinary disk from the inner circumstellar disks (Lin & Papaloizou 1993; Artymowicz and Lubow 1994). In spectroscopic binaries, with orbital separations less than ~ 1 AU, the affected regions of the disk are those that produce the near- and mid-infrared excesses in CTTS. Therefore, one signature of a disk gap should be a deficit of dust emission at these wavelengths, a dip in the spectral energy distribution (SED), compared to a power-law SED expected for a disk around a single CTTS. Indeed, the SEDs of some CTTS spectroscopic binaries show dips in the infrared, and the inferred gap sizes for these systems are consistent with those predicted by dynamical theory (Mathieu, Adams, & Latham 1991; Jensen & Mathieu 1997).

A giant planet ($\geq 1\ M_{JUP}$) forming within a protoplanetary accretion disk is also expected to open up a gap at its orbital radius. The formation of a gap could be significant in terminating or limiting the growth of massive planets (Lin & Papaloizou 1993; Bryden et al. 1999; Kley 1999; Lubow et al. 1999). The tidal interaction between a giant protoplanet and its protoplanetary disk is also thought to produce orbital migration (Goldreich & Tremaine 1980; Hourigan & Ward 1984; Lin & Papaloizou 1986), which could vastly affect the final orbital distribution of massive planets and their rate of survivability, if massive planets migrate into the central star (Ward 1997). The importance of these predicted physical processes makes it highly desirable to observationally confirm the existence of disk gaps and to obtain measurements (such as the radial location and widths of gaps, and the amount of material within them) that can be compared to theory.

Earlier theoretical work suggested that a dynamically cleared gap in a binary system would isolate the inner circumstellar disk(s) from the outer circumbinary disk. Without a means of replenishment, the circumstellar disk(s) would deplete on a viscous timescale, ending accretion onto the stellar surface(s). This timescale is relatively short for the orbital separations of spectroscopic binary CTTS, and the evidence for continuing high rates of accretion onto these stars (similar to accretion rates in single or wide binary CTTS) is in apparent contradiction to the idea that a gap will cut off accretion (Mathieu et al. 2000). More recent numerical simulations (Artymowicz & Lubow 1996), however, show that it may be possible for gas from the circumbinary disk to cross disk gaps in the form of accretion streams, feeding matter onto the binary system with an average accretion rate similar to the unimpeded rate for a disk about a single star.

Dramatic evidence for this idea comes from the eccentric (e=0.556) double-lined spectroscopic binary DQ Tau. Photometric and spectroscopic monitoring of DQ Tau shows periodic brightenings with corresponding increases in optical veiling and emission line strengths, with a period equal to the binary orbital period of 15.8 days (Mathieu et al. 1997; Basri, Johns-Krull, & Mathieu 1997). This is the type of modulated, or pulsed, accretion predicted by Artymowicz & Lubow (1996) for an eccentric, equal mass system such as DQ Tau. In addition, the maximum accretion rate is observed to occur at periastron, as predicted.

A second apparent contradiction to the idea of disk clearing in spectroscopic binary CTTS is the lack of a dip in the SEDs for some systems (Jensen & Mathieu 1997). DQ Tau, along with UZ Tau E (P=19.1 days) and AK Sco (P=13.6 days), have power-law SEDs in the infrared that are indistinguishable from other CTTS. Mathieu et al. (1997) and Jensen & Mathieu (1997) suggest that near- to mid-infrared emission, sufficient to fill in the expected dip in the SED, is produced by small amounts of optically thin dust residing within the gap. This is not necessarily surprising; for example, a reduction in the column density of the minimum mass solar nebula (~1500 g cm$^{-2}$ at 1 AU) by a factor of $10^6$ still leaves sufficient material to achieve a continuum optical depth of ~ 0.01 at 5 µm for a standard dust-to-gas ratio. Strong silicate emission observed from some of the systems is also consistent with the presence of hot optically thin dust (Mathieu et al. 1991; Jensen & Mathieu 1997). Thus, while dynamically cleared gaps may exist, they may not be completely devoid of material.

In general, disk gaps may contain some residual gas and dust, either because dynamical clearing is not entirely efficient, or accretion streams that cross the gap supply a low surface-density of material. The gas in these optically-thin regions of an otherwise optically-thick accretion disk should be observable via emission lines. The measurement of emission lines from gap material would be highly advantageous, because the kinematics of the profiles observed at high-spectral resolution can pinpoint the radii of the gas (for known inclination and stellar mass), and the measurement of several appropriately chosen transitions can be used to determine the gas temperatures and column densities. Because gas temperatures within ~ 1 AU of the star will be in the range of a couple hundred to a couple thousand degrees, rotational and ro-vibrational transitions of molecules in the infrared are a logical choice for tracers of the gas.

In order to explore this idea, we have obtained high-resolution infrared spectra of spectroscopic binary CTTS to search for emission from the CO fundamental transitions near 4.7 μm. This paper presents the detection and analysis of CO $v$=1–0 and $v$=2–1 emission from the spectroscopic binary DQ Tau.

2. OBSERVATIONS

High-resolution spectra of DQ Tau were obtained at the NASA Infrared Telescope Facility using CSHELL, the facility cryogenic infrared echelle spectrograph (Tokunaga et al. 1990). CSHELL uses circular variable filters to isolate a single echelle order on the 256x256 SBRC InSb array, giving a spectral coverage of ~ 750 km s$^{-1}$. The central wavelength was set to either the R(3) or the P(18) line of the $v$=1–0 band, at 4.63 and 4.83 μm, respectively. A log of the observed spectra is given in Table 1, which lists the orbital phase of DQ Tau and the S/N obtained on the continuum. For each spectrum of DQ Tau, the hot star HR 1412 was observed at approximately the same airmass. All of the spectra were obtained with a 1.0" slit, giving a nominal resolution of ~ 14 km s$^{-1}$.

The telescope was nodded alternately between two positions along the slit. During data reduction, these slit images were differenced to subtract the background emission, which is the dominating noise source at 4.7 µm. The difference images were then averaged and a spectrum extracted for each slit position. Separate dispersion solutions were determined for each of the slit positions before combining them into one spectrum. Telluric absorption lines in the hot star spectrum were used for the wavelength calibration. The hot star spectrum was divided into the DQ Tau spectrum to correct for telluric absorption lines.

The DQ Tau spectra were flux calibrated by obtaining a short exposure in a 4" slit of both DQ Tau and HR 1412, and then using the ratio of these to scale the longer exposure spectrum taken with the 1" slit. Given a K magnitude of 2.92, and adopting K–M = 0.03, the flux density for HR 1412 is $1.48 \times 10^{-12}$ W m$^{-2}$ µm$^{-1}$ at 4.80 µm. A blackbody of 8000 K was used to set the flux density at other wavelengths. The relative flux calibration among the DQ Tau spectra is estimated to be 10 % or better. Due to the arrival of clouds, the spectrum on 1995 January 25 was not flux calibrated. The velocity calibration uncertainty for the individual spectra is estimated to be 2–3 km s$^{-1}$.

## 3. RESULTS

The spectrum of DQ Tau shows CO emission lines from both the $v=1$ and $v=2$ vibrational levels. Spectra of the R(3) $v=1$–0 line (hereafter R(3) ), observed at three different epochs, are compared in Figure 1. The higher SNR of the 1998 R(3) spectrum clearly revealed the weaker $v=2$–1 R(11) and R(10) lines (Fig. 2). A spectrum of the $v=1$–0 P(18) line (hereafter P(18)) was also obtained in 1998 (Fig. 2). Portions of the spectra that fall within the cores of strong telluric lines are not plotted in Figures 1 and 2. The measured heliocentric velocity, line width, equivalent width, and line flux for the emission lines are listed in Table 2.

The centroid velocity of the R(3) line is the same at all three observed epochs and, within the uncertainties, is equal to the center-of-mass radial velocity of 22.4 km s$^{-1}$ for DQ Tau (Mathieu et al. 1997). The average of the line widths (FWHM) is 67 km s$^{-1}$, and the $v=2$–1 lines appear to be somewhat broader than the $v=1$–0 lines. The predicted velocities of the stellar components of DQ Tau, and the maximum stellar velocity separation of 55 km s$^{-1}$, are indicated in Figure 1. The SNR of the spectra are too low to determine if there is any variation of the line profile with orbital phase, but the similarity of the line width and centroid velocity at different phases suggests that the emission is not associated with the individual stars or their circumstellar disks. Better time coverage with higher signal-to-noise is required to investigate this in more detail.

Both the integrated R(3) line flux and continuum flux changed by 30 % between the 1996 and 1998 spectra, suggesting a related physical origin. When the broadband 4.8 μm flux measurement from 1996 October 31 UT is included (Mathieu et al. 1997), the M-band continuum of DQ Tau shows a range of a factor of two. The R(3) equivalent width did not vary measurably between 1996 and 1998; in the uncalibrated 1995 spectrum the equivalent width may be slightly (3-σ) smaller.

The measured line fluxes can be used to determine the excitation temperature of the CO gas by assuming that the emission is optically thin. For the 1998 data, Figure 3 shows the natural log of $F_{ul}/\nu g_u A_{ul}$ plotted against the upper energy level, where $F_{ul}$ is the line flux, $\nu$ the line frequency, $g_u$ the statistical weight of the upper level, and $A_{ul}$ the transition probability. Because $F_{ul}/\nu g_u A_{ul}$ is proportional to the population of the upper level divided by the statistical weight, the inverse of the slope in this plot is the negative of the excitation temperature. The relative flux calibration uncertainty between the R(3) and P(18) spectra has been incorporated into the P(18) point, and blending with weaker CO emission lines (see Section 4) has been ignored. A fit to all four data points yields T = 1252 ± 49 K; this is not significantly different from a fit to only the two $v$=1–0 lines, which gives T = 1143 ± 180 K. The consistency between the rotational and vibrational temperatures suggests that the rotational and vibrational levels are close to being in LTE, at least up to the $v$=2 level.

An average temperature of ~ 1200 K is suggestive of circumstellar material in the vicinity of the binary. This is similar to the temperature (1000 K) of the optically thin dust postulated by Mathieu et al. (1997) to occupy the dynamically cleared gap in DQ Tau. In addition, the observed widths of the R(3) lines are comparable to the 55 km s$^{-1}$ maximum projected velocity separation of the stars, suggesting that at least some of the emission comes from material within the expected gap.

In the next section, we calculate disk emission models for the CO lines in order to estimate the parameters required to reproduce the observed emission.

## 4. DISK MODELING

To model the gas emission from the disk, we envision that the binary system lies at the center of an optically thick circumbinary disk (Mathieu et al. 1997). The binary has cleared a hole in this circumbinary disk, and in this hole resides optically thin dust and gas. Surrounding each star may be very small, optically thick circumstellar disks.

The disk modeling procedure was similar to that described in Najita et al. (1996), except that here the line emission region was modeled as a single layer and dust opacity was included, using the opacities of Henning & Stognienko (1996) for inhomogeneous dust aggregates. The CO abundance relative to hydrogen was calculated assuming chemical equilibrium, and LTE was assumed. The line list for the model spectra included all $^{12}$CO lines and $^{13}$CO lines. The $v$=2–1 R(10) and R(11) lines have small contributions from $v$=1–0 $^{13}$CO lines, the R(3) has a small contribution from the $v$=3-2 R(19) line, and the P(18) line is blended with the $v$=2–1 P(12) line. We took a combined binary mass for DQ Tau of 1.3 $M_\odot$, a system inclination of 23º (Mathieu et al. 1997), a distance of 140 pc, and zero extinction at 4.7 μm. The uncertainty in the effective temperature for the stars, and comparisons to different sets of evolutionary tracks, indicate an uncertainty in the mass of ~ 25 % and a 11 % uncertainty in sin $i$. Finally, a constant continuum was added to the model to match the observed continuum level; the observed continuum is roughly an order of magnitude greater than the continuum derived from the CO-to-dust emissivity ratio of our model. This continuum is purely additive and does not affect the formation of or radiative transfer within the observed CO lines. Its significance is discussed in Section 5.

We are not in a position to model the two-dimensional spatial distribution of gas in a binary system. Hence we model the system with an axisymmetric, Keplerian model with the combined mass of the stars. This model will be reasonably accurate near the inner edge of the circumbinary disk. In the vicinity of the binary orbit we can only claim that the Keplerian model approximates the scale of the orbital motions. In any case, the SNR of the line profiles is inadequate to distinguish differences from Keplerian rotation.

Model spectra were compared to the higher quality 1998 data. The free parameters in the model which can be varied to fit the observed lines are the radial variation of the gas temperature and surface density, and the inner and outer radius of the gap. The relative line strengths are largely determined by the gas temperature, while the absolute line fluxes are determined by the combination of gas temperature, surface density, and emitting area. The model line profile and line width are set by the inner and outer radii and the radial variation of the emission surface brightness (temperature and surface density).

The gas temperature and disk surface density were first set to be constant with radius. Modeling of the R-branch $v=1$–$0$ and $v=2$–$1$ lines required a gas temperature of 1150 K, a surface density of $4 \times 10^{-4}$ g cm$^{-2}$, and an outer radius of ~ 0.5 AU. With these parameters, however, it was not possible to simultaneously fit the width of both the $v=1$–$0$ R(3) line and the wider, higher excitation $v=2$–$1$ lines. This suggests that a temperature gradient would improve the agreement by preferentially increasing the emission of the higher excitation $v=2$–$1$ lines with respect to the $v=1$–$0$ lines from smaller radii where the projected velocities are larger. (A gradient in the surface density alone would not alter the relative widths of the lines). We allowed for a temperature gradient of the form $T(r) \propto r^{-0.5}$, again with a constant surface density. With this temperature gradient a reasonable fit to the spectrum was found for an outer radius of 0.5 AU, a temperature of 770 K at the outer radius, and a surface density of $5 \times 10^{-4}$ g cm$^{-2}$. This model is overplotted on the data for the R-branch lines in Figure 4(a), with the velocity of the model spectrum set to the center-of-mass velocity for DQ Tau.

The modeling uncertainty in the outer radius of 0.5 AU is about 0.1 AU. An outer radius greater than 0.6 AU, or smaller than 0.4 AU, produces model line profiles that are too narrow, or too wide, respectively. The uncertainties in the system inclination and stellar mass introduce a similar uncertainty in the outer radius of 0.08 AU. Due to the low SNR in the line wings, the inner radius is not tightly constrained, but values greater than ~ 0.1 AU produce profiles that are unacceptably narrow, particularly in the $v$=2–1 lines and in the wings of the R(3) line. Thus, the model indicates that the optically thin region extends to within close proximity of the stars. Most of the line flux in this model is produced at temperatures between about 800 and 1700 K, consistent with the constant temperature model. For these temperatures and column density, the R(3) line is marginally optically thin, with an optical depth ~ 0.3, while the optical depth in the two $v$=2–1 lines is ~ 0.06. The small continuum optical depth in the gap due to dust, about 0.002 at 4.8 μm, has no effect on the line formation (neither would the larger dust column density inferred by Mathieu et al 1997). Instead of modeling the P(18) and R-branch spectra simultaneously, the model derived from the R-branch lines was applied to the P(18) line to check for consistency. The model somewhat over-predicts the observed P(18) flux, but the difference is within the relative calibration uncertainty. The model spectrum is over-plotted on the P(18) line in Figure 4(b), where the observed spectral flux has been increased by 8 %. It should also be noted that the P(18) line was observed at a different date and phase from the R-branch lines (see Table 1).

The total mass involved, obtained by integrating the gas surface density over the area of the gap, is quite small, $4 \times 10^{-11}$ $M_\odot$ or only $10^{-5}$ $M_\oplus$. This mass is similar to, but roughly one order of magnitude smaller than, the $5 \times 10^{-10}$ $M_\odot$ of material that was required by Mathieu et al. (1997) to explain the apparent lack of a mid-infrared dip in DQ Tau's SED. The difference in mass could be due to a larger dust-to-gas ratio than the ISM, a smaller CO/H abundance ratio than the chemical equilibrium value, non-LTE excitation of CO, or some combination of these factors. As previously noted, the CO gas temperature is nearly the same as the dust temperature required by Mathieu et al.

## 5. DISCUSSION

Because circumbinary material must cross the gap in order to feed the on-going accretion onto the stars, the discovery of residual gas and dust distributed within the gap is important. A critical question is whether the observed material in the gap is consistent with the observed accretion rates. Estimates of the accretion rate for DQ Tau range from $2 \times 10^{-8}$ $M_\odot$ yr$^{-1}$ (Valenti, Basri & Johns 1993) and $5 \times 10^{-8}$ $M_\odot$ yr$^{-1}$ (Hartigan, Edwards & Ghandour 1995) to as low as $6 \times 10^{-10}$ $M_\odot$ yr$^{-1}$ (Gullbring et al. 1998). The latter is probably unrepresentative of the average accretion rate, because Gullbring et al. measured an upper limit on the optical veiling of 0.06. Based on monitoring by Basri et al. (1997), the veiling has a typical value of 0.35, occasionally drops to zero, and reaches peak values of 1.5 near periastron. Taking an accretion rate of ~ $10^{-8}$ $M_\odot$ $yr^{-1}$ and a dynamical time for material flowing through the gap comparable to the binary orbital period, the mass in the gap required to feed the accretion rate is 5 x $10^{-10}$ $M_\odot$.

This mass is the same as the rough estimate for the mass in the gap derived from the near-infrared continuum emission by Mathieu et al. (1997). However, it is an order of magnitude larger than the total mass of 4 x $10^{-11}$ $M_\odot$ inferred from the CO emission for material in the gap.

Put in a different way, the theoretical dust emission associated with the observed CO line strengths is an order of magnitude less than observed. An additional source for the observed near-infrared continuum emission is not evident. Unless the inner edge of the circumbinary disk is unexpectedly warm, it cannot provide the observed emission (Mathieu et al. 1997, Fig. 6). Dynamically stable circumstellar disks are much too small (less than 4 stellar radii) to provide the required emission.

Unfortunately, both the dust providing the continuum emission and the CO providing the line emission are expected to be trace components of protostellar disks, which are predominantly molecular hydrogen. Thus the derived total masses are sensitive to uncertainties in the relative abundances of dust and CO to hydrogen. The masses derived from the measured accretion rate, the continuum flux, and the CO could be reconciled at 5 x $10^{-10}$ $M_\odot$ if the CO emissivity of the disk material is an order of magnitude lower than in our model. One likely possibility is subthermal excitation of the CO vibrational levels, given the temperature (T < 1500 K) and low density (n < $10^{10}$ $cm^{-3}$) in the gap region. Modeling that includes non-LTE excitation (see Najita et al. 1996) should clarify this issue. The CO abundance could also be substantially lower than our assumed chemical equilibrium value, either because CO is not the main carbon-bearing molecule, or CO in the gap is photodissociated by UV radiation from the central star and shock. We note that a lower CO emissivity of the material in the gap would remove the need for the arbitrary addition of continuum flux in our modeling (Section 4). This would be achieved without changing the model

CO line profiles, since both the dust and CO are optically thin. This linkage of the continuum and line emission would also more easily explain the correlated variation in the continuum and line brightnesses (Section 3).

Regardless of the value for the total mass, the CO line ratios allow us to place constraints on the spatial distribution of the material. Since we have detections in several lines, the relative optical depths can constrain the emitting surface area. Within the context of our disk model, we examined this by introducing a surface filling factor for the gas. As the emitting surface area is decreased, and the CO mass held constant, the R(3) line quickly becomes optically thick, and further decreases in the surface area lower the R(3) emission line flux while the optically thin $v$=2–1 line fluxes remain constant. Alternatively, the mass can be increased in an attempt to maintain a constant R(3) flux, but the faster growing $v$=2–1 growing lines then become too strong. Quantitatively, we find that matching the observed lines ratios is a clear problem for filling factors less than 30 %. For filling factors smaller than 10 %, sufficient flux cannot be produced in the R(3) line, regardless of the CO mass.

In the simulations of Artymowicz & Lubow (1996), accretion takes the form of streams of material crossing a nearly empty gap. The accretion onto the stars is predicted to be highly pulsed for the case of an eccentric equal-mass binary like DQ Tau, as observed. If such streams exist in the DQ Tau system, our analysis suggests that they have a filling factor greater than 30 %. The presence of continuous low-level accretion activity at all orbital phases indicates that at least near the stars there is a less ordered distribution of material (Mathieu et al. 1997; Basri et al. 1997). A distribution of material that includes both poorly organized accretion streams and gas dispersed throughout the gap may be consistent with the CO observations.

Of course, the gas motions of infalling accretion streams are different than those of the rotating disk that we have modeled. While actual modeling of line profiles from accretion streams would be needed for quantitative comparisons, the expected infall velocities are of the right magnitude to plausibly produce the observed line widths. The important point for the above discussion is that the arguments concerning the total CO mass, line optical depths, and emitting area are not strongly dependent on the assumed velocity field.

It is important to emphasize that the DQ Tau system is characterized by substantial variability on numerous timescales. The observations by Mathieu et al. (1997) and Basri et al. (1997) show that the accretion onto the stars is highly variable both within an orbit and from orbit to orbit. Similarly, in the Artymowicz & Lubow (1996) models, the accretion stream is variable over an orbital period as material from the inner edge of the circumbinary disk is perturbed near apastron and over the next half an orbital period falls within the stars' Roche lobes. Hence, both the flux and line profile of CO emission from an accretion stream is expected to be variable. The data modeled here reflect only one orbital phase (0.37). Higher signal-to-noise data at several orbital phases should substantially clarify the evolution of the spatial distribution of material in the gap.

Finally, the fact that the SED for DQ Tau can be fit by an optically thick accretion disk without a gap does suggest another possibility for the origin of the observed CO emission: that it arises from an optically-thin temperature inversion layer in a continuous optically-thick disk (as in Calvet et al. 1991; D'Alessio et al. 1998). A detailed calculation and predicted spectrum for the case of DQ Tau would be required to show that a disk inversion layer would have the necessary temperatures and column densities to produce the observed CO emission. Based on dynamical grounds, however, it seems unlikely that an optically-thick disk could exist at the binary orbital separation without tidal disruption. In the more general case, discrimination between an optically-thin disk gap and a disk temperature inversion could be made on the radial distribution of the emission. A gap will cover a distinct range in radii, while an inversion layer should show a gradual decrease in emissivity with increasing radius with no particular radial limits. The relatively abrupt emission cut-off at 0.5 AU for DQ Tau appears to favor emission from a gap or inner hole.

## 6. CONCLUSIONS

Our measurement of CO fundamental emission from DQ Tau is the first detection of these lines in a low-mass T Tauri star. The widths of the emission lines and our disk models for the profiles are consistent with gas in orbital motion within an outer radius of 0.5 ± 0.1 AU from the center-of-mass. This radius is consistent with a dynamical gap with ~ 0.4 AU outer radius that is expected for the DQ Tau system (Mathieu et al 1997) based on the analytic calculations and smoothed particle hydrodynamic simulations of Artymowicz & Lubow (1994; 1996) for eccentric binary systems. The temperature derived for the gas is reasonable for material at these distances from the star and is the same as that of the optically-thin dust proposed by Mathieu et al. (1997) to reside within the gap. The mass derived from the CO emission (4 x $10^{-11}$ $M_{\odot}$) is an order of magnitude less than that indicated by the near- and mid-infrared continuum emission and the measured accretion rates (5 x $10^{-10}$ $M_{\odot}$). This discrepancy may indicate that the CO emissivity of the gap material is smaller than as modeled.

Evidently, dynamical clearing has not been completely efficient in the DQ Tau binary. Equally important, the newly discovered material in the gap strongly suggests that the ongoing accretion at the stellar surfaces is fed by material flowing inward from the circumbinary disk. The presence of material in the gap is intriguing in the context of previously suggested accretion streams from the circumbinary disk to the stars. The CO line emission can place restrictions on the area of such streams but does not directly indicate their presence.

This result, and the fact that the SED for DQ Tau does not suggest nor require the presence of a disk gap, suggests that line emission from residual gas within gaps may be a more robust diagnostic of disk gaps than SEDs. An analysis of CO fundamental emission from other spectroscopic binary CTTS, and more definitive modeling based on higher signal-to-noise ratio line profiles, will determine the degree to which CO fundamental lines (or other gas diagnostics) can be used as tracers of the physical properties of disk gaps. If further work proves this to be a positive approach, then this technique can provide important constraints on the physics of gap clearing in disks, with significant potential for testing theories of giant planet formation; it also holds the exciting potential to indirectly detect the presence of protoplanets within their protoplanetary disks (Carr & Najita 1998).

JSC acknowledges support from a NASA Origins of Solar Systems grant and the Office of Naval Research. RDM was supported by NSF grant AST-941715.**REFERENCES**

Artymowicz, P. & Lubow, S. H. 1994, ApJ, 421, 651

Artymowicz, P. & Lubow, S. H. 1996, ApJL, 467, L77

Basri, G., Johns-Krull, C. M., & Mathieu, R. D. 1997, A. J., 114, 781

Bryden, G., Chen, X., Lin, D. N. C., Nelson, R. P., Papaloizou, J. C. B. 1999, ApJ, 514, 344

Calvet, N., Patiño, A., Magris C., G., & D'Alessio, P. 1991, ApJ,

Carr, J. S., & Najita, J. 1998, in ASP Conf. Ser. 133, Science With The NGST, e.d. E. P. Smith
    & A. Koratkar (San Francisco: ASP), 163

D'Alessio, P., Cantö, J., Calvet, N., & Lizano, S. 1998, ApJ, 495, 385

Goldreich, P. & Tremaine, S. 1980, ApJ, 241, 425

Gullbring, E., Hartmann, L., Briceno, C., & Calvet, N. 1998, ApJ, 492, 323

Hartigan, P., Edwards, S., & Ghandour, L. 1995, ApJ, 452, 736

Henning, Th. & Stognienko, R. 1996, A&A, 311, 291

Hourigan, K., & Ward, W. R. 1984, Icarus, 60, 29

Jensen, E. L. N. & Mathieu, R. D. 1997, A. J., 114, 301

Kley, W. 1999, MNRAS, 303, 696

Lin, D. N. C. & Papaloizou, J. 1986, ApJ, 309, 846

Lin, D. N. C. & Papaloizou, J. C. B. 1993, in Protostars and Planets III, ed. E. H. Levy & J. I.
    Lunine, (Tucson: Univ. of Arizona Press), 749

Lubow, S. H., Seibert, M., & Artymowicz, P. 1999, ApJ, 526, 1001

Mathieu, R. D., Adams, F. C., & Latham, D. W. 1991, AJ, 101, 2184

Mathieu, R.D., Ghez, A., Jensen, E.L.N. & Simon, M., 2000, in Protostars and Planets IV,, eds. V.

Mannings, A. Boss & Russell, (Tucson: Univ. of Arizona Press), 703.
15

Mathieu, R. D., Stassun, K., Basri, G., Jensen, E. L. N., Johns-Krull, C. M., Valenti, J. A., & Hartmann, L. W. 1997, A. J., 113, 1841

Najita, J., Carr, J. S., Glassgold, A. E., Shu, F. H. & Tokunaga, A. T. 1996, ApJ, 462, 919

Tokunaga, A. T., Toomey, D. W., Carr, J. S., Hall, D.N.B. & Epps, H. W. 1990, SPIE, 1235, 131

Valenti, J.A., Basri, G., & Johns, C.M. 1993, AJ, 106, 2024

Ward, W. R.. 1997, ApJ, 482, L211

**FIGURE CAPTIONS**

Figure 1. Spectra of the $v=1–0$ R(3) line observed at three epochs, as labeled. The long dashed line marks the center-of-mass radial velocity for the DQ Tau system, the short solid lines mark the radial velocity of the primary, and the short dotted lines mark the radial velocity of the secondary. The horizontal bar in the top panel shows the maximum stellar velocity separation. The gap in the spectra is the core of the telluric CO absorption.

Figure 2. top) The 1998 January spectrum of the R-branch CO lines in DQ Tau, and bottom) the $v=1–0$ P(18) line. The upper x-axis in each panel is the heliocentric radial velocity. The gaps in the spectra are regions of strong telluric absorption.

Figure 3. Excitation temperature for CO lines. Plotted against the energy of the upper level is the natural log of the line flux divided by the product of the line frequency, statistical weight of the upper level, and transition probability. The data points for the $v=1–0$ R(3) and P(18) lines, and the $v=2–1$ lines (R(10) and R(11)) are indicated. The solid line is a fit to all four points, yielding T = $1252 \pm 49$ K. The dotted line is fit to only the $v=1–0$ R(3) and P(18) lines, yielding T = $1143 \pm 180$ K.

Figure 4. a) Disk-gap model with radial temperature gradient (see text) over-plotted on the 1998 spectrum of the R-branch lines. The disk-model surface density is $5 \times 10^{-4}$ g cm$^{-2}$, the outer radius 0.5 AU, the temperature is $770(R/0.5\ \mathrm{AU})^{-0.5}$, and the inner radius $\leq 0.1$ AU. b) The same model compared to the spectrum of the $v=1–0$ P(18) line. For comparison purposes, the flux of the P(18) spectrum has been increased by 8 %, a value within the relative calibration uncertainty.

TABLE 1

DQ Tau Observations

| CO Line | UT Date | UT Time | Phase | Cont. SNR |
|---|---|---|---|---|
| 1–0 R(3) | 1995 Jan 25 | 08:30 | 0.13 | 5.5 |
| 1–0 R(3) | 1996 Jan 16 | 08:57 | 0.64 | 7.5 |
| 1–0 R(3) | 1998 Jan 08 | 06:59 | 0.37 | 15 |
| 1–0 P(18) | 1998 Jan 10 | 07:52 | 0.50 | 7 |

TABLE 2

CO Line Measurements

| CO Line | UT Date | $V_{hel}$ (km s$^{-1}$) | FWHM (km s$^{-1}$) | EW (Å) | Line Flux (10$^{-17}$ W m$^{-2}$) |
|---|---|---|---|---|---|
| 1–0 R(3) | 1995 Jan 25 | +27 ± 6 | 54 ± 6 | 4.5 ± 0.3 | … |
| 1–0 R(3) | 1996 Jan 16 | +22 ± 6 | 69 ± 8 | 6.0 ± 0.4 | 2.8 ± 0.2 |
| 1–0 R(3) | 1998 Jan 08 | +24 ± 6 | 58 ± 6 | 6.1 ± 0.4 | 3.6 ± 0.2 |
| 2–1 R(11) | 1998 Jan 08 | +28 + 4 | 69 ± 3 | 2.1 ± 0.3 | 1.3 ± 0.2 |
| 2–1 R(10) | 1998 Jan 08 | +20 ± 4 | 83 ± 4 | 2.3 ± 0.3 | 1.5 ± 0.2 |
| 1–0 P(18) | 1998 Jan 10 | +21 ± 5 | 68 ± 4 | 9.4 ± 0.4 | 6.9 ± 0.3 |

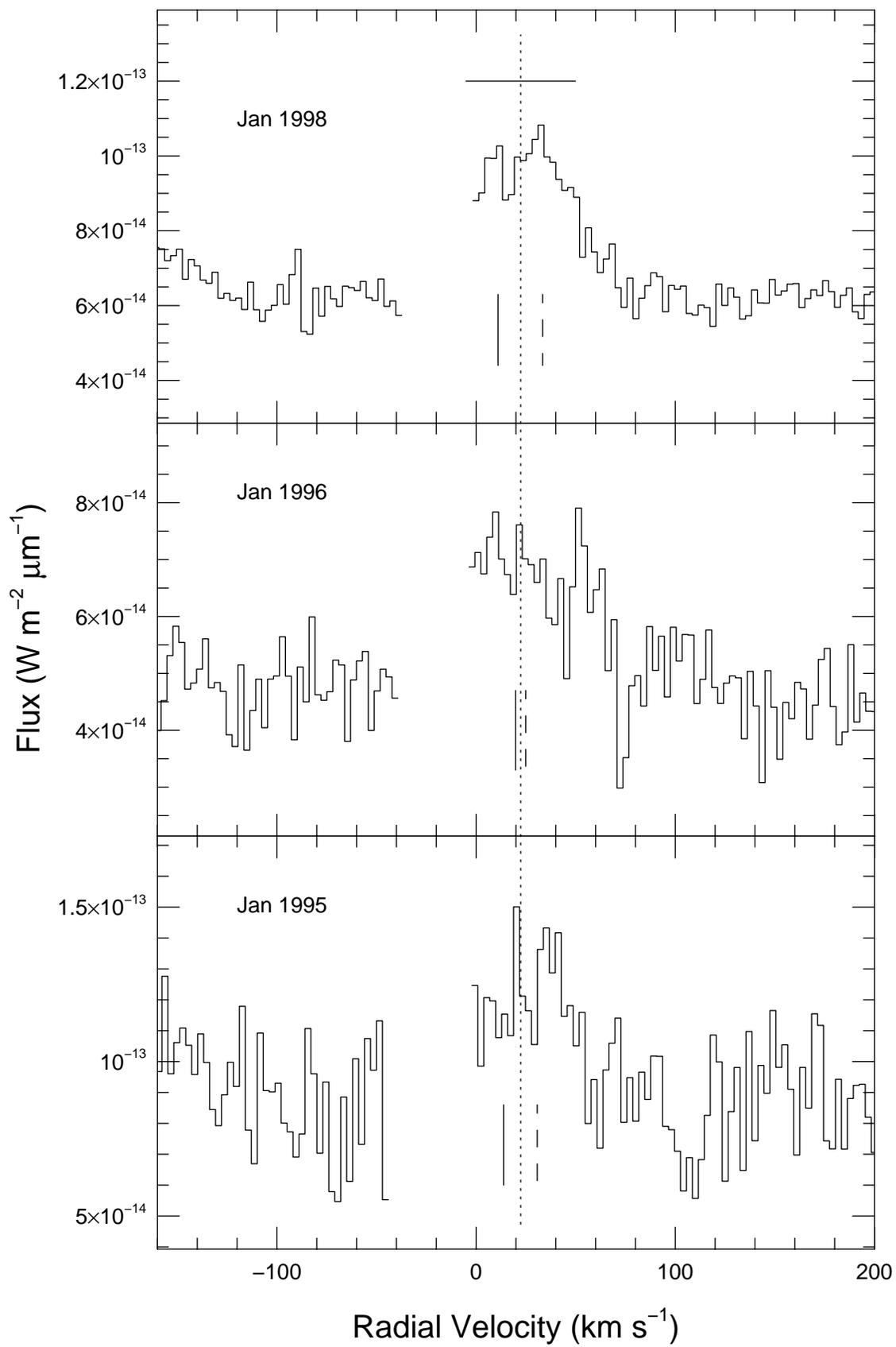

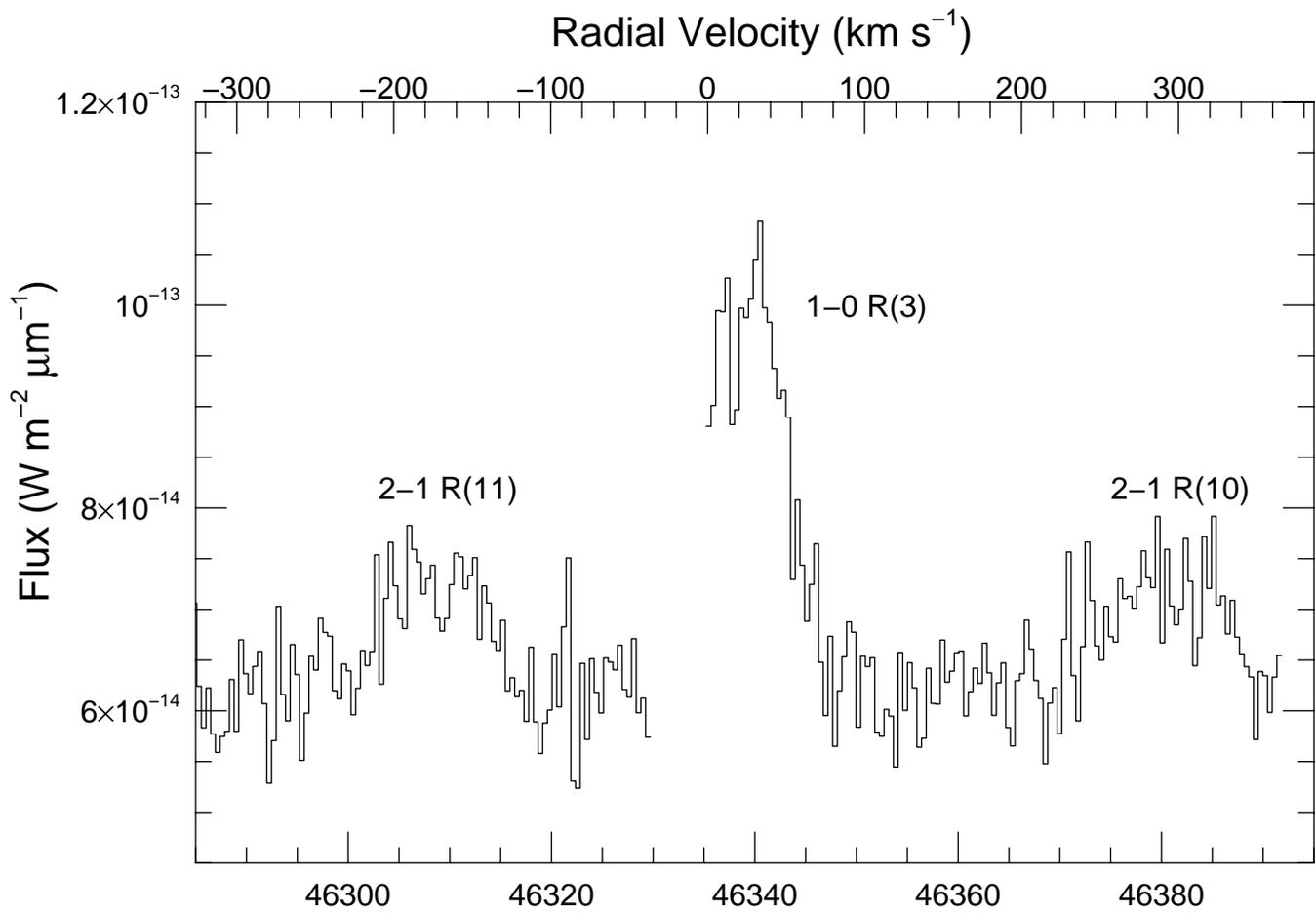
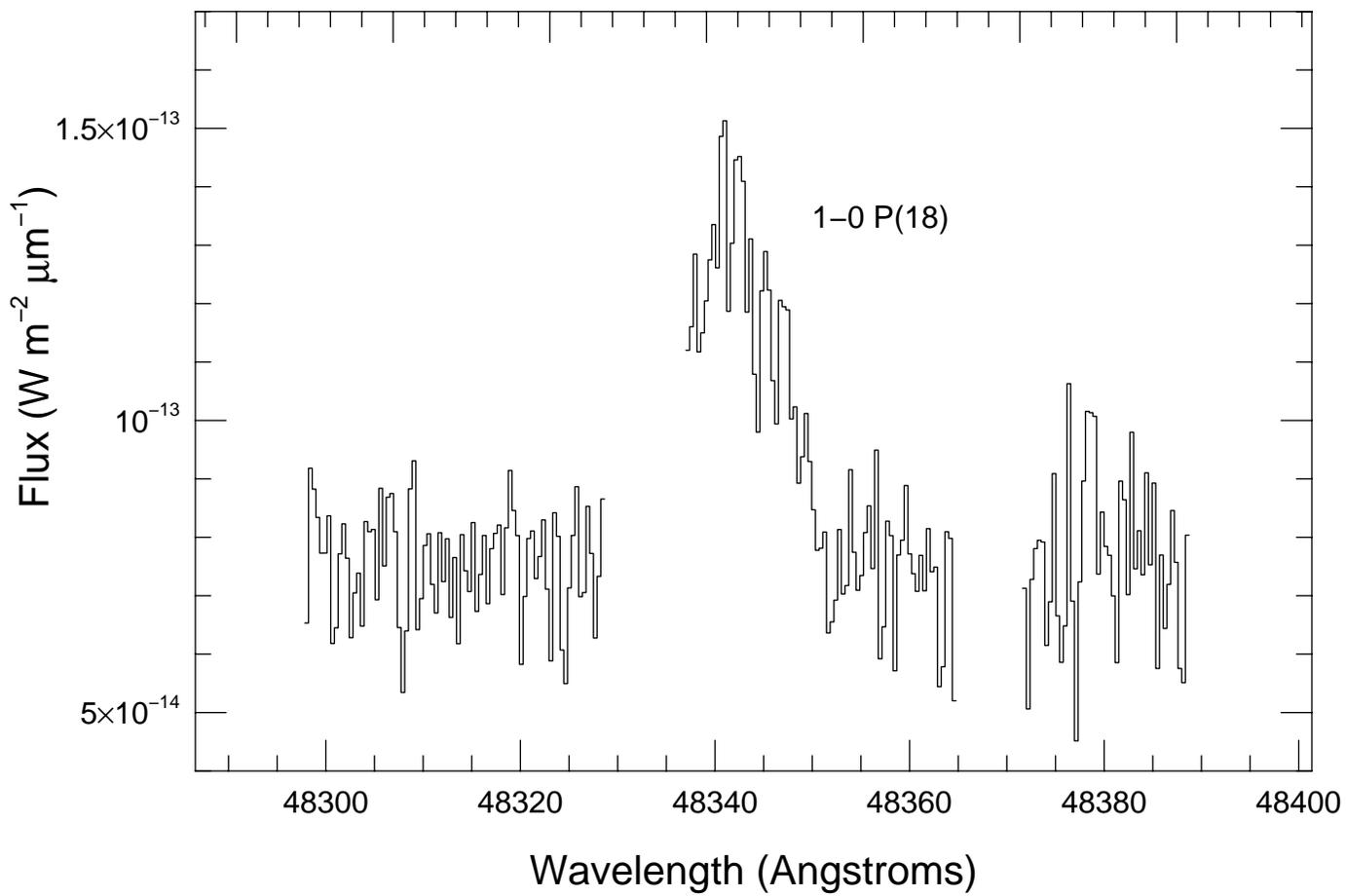

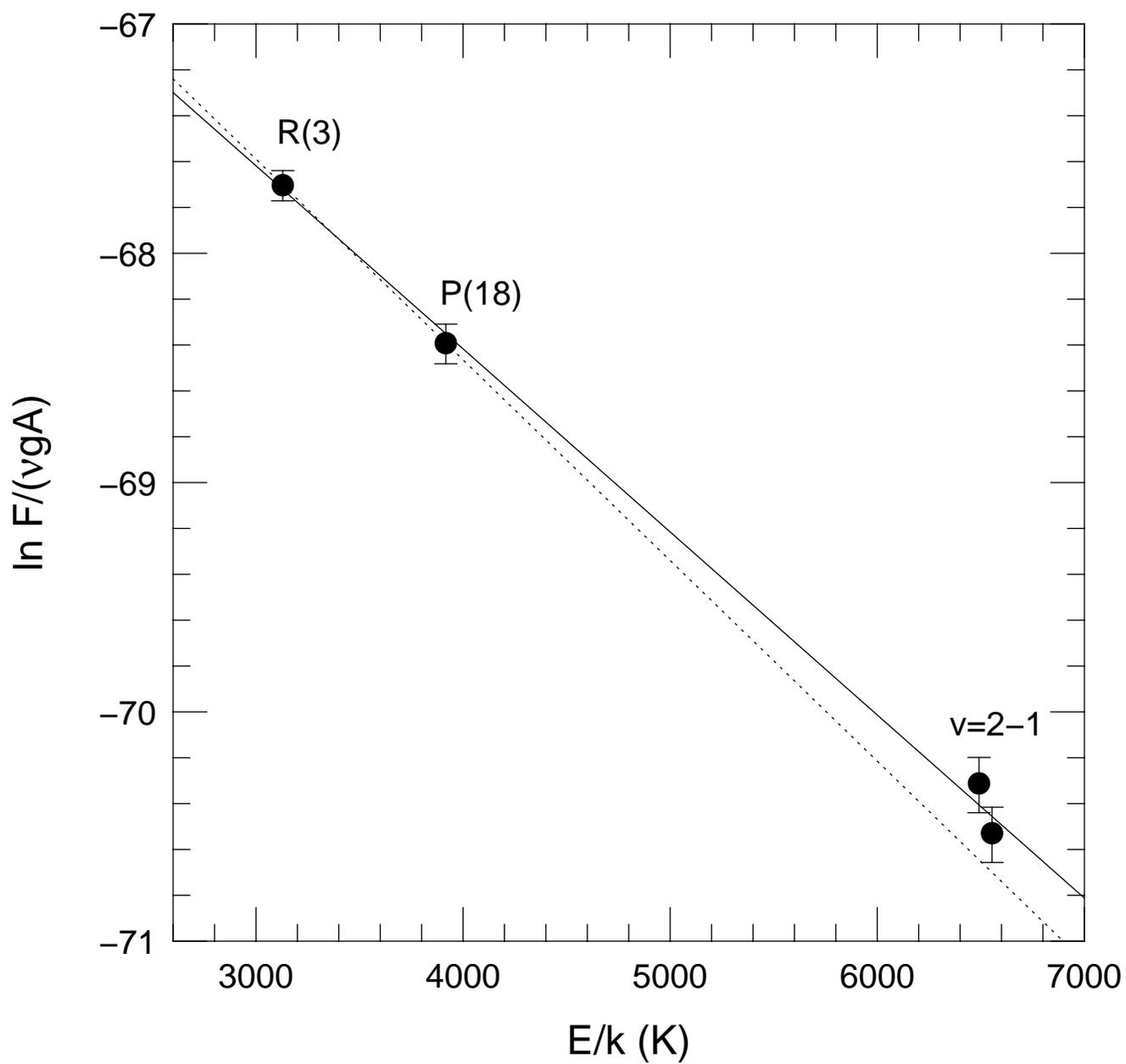

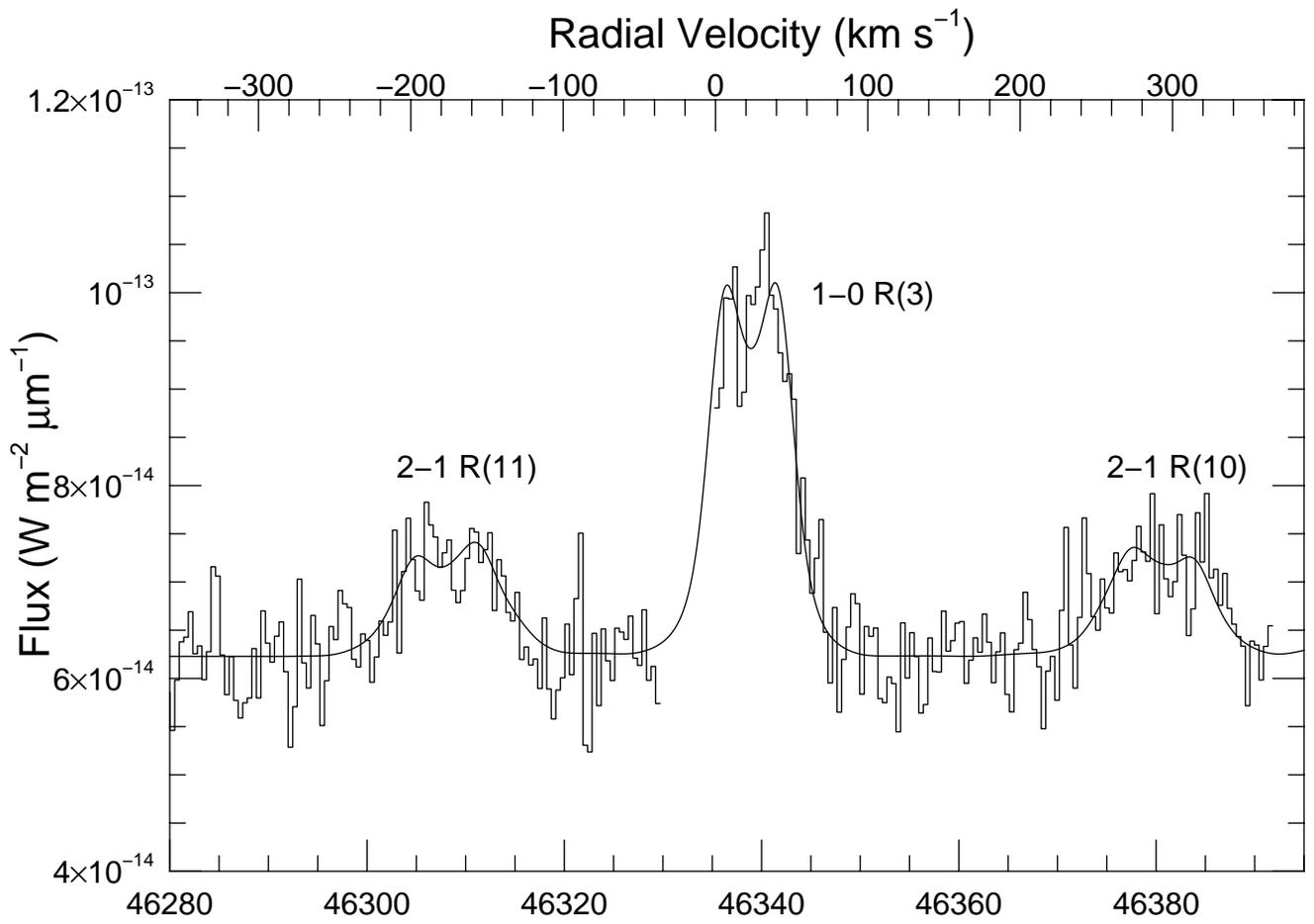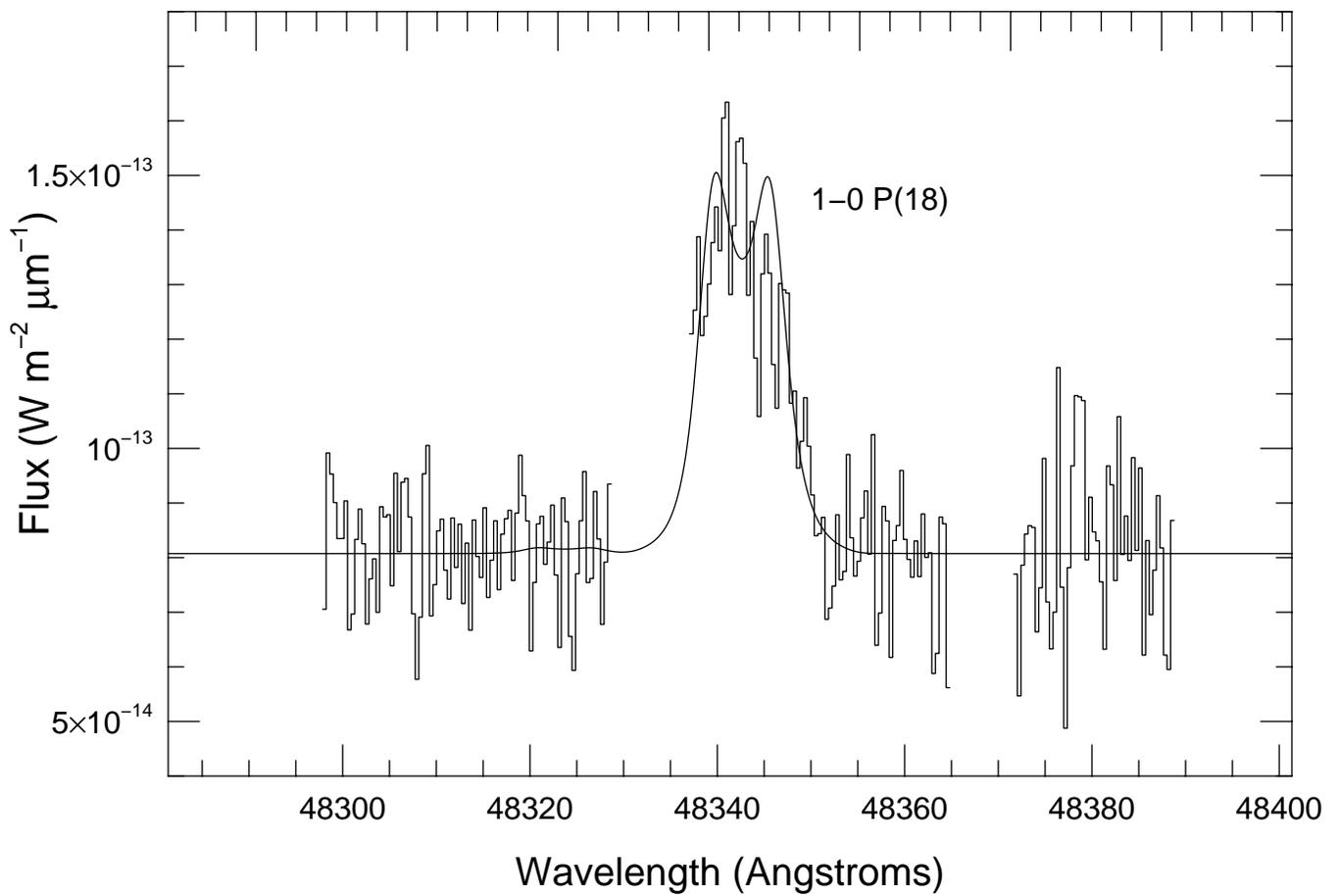